\documentclass[%
reprint,
superscriptaddress,
amsmath,
amssymb,
aps,
prl,
longbibliography
]{revtex4-2}

\usepackage{amsmath}
\usepackage{amsfonts}
\usepackage{amssymb}
\usepackage{array}
\usepackage{bbm}
\usepackage{bm}
\usepackage{bbold}
\usepackage{braket}
\usepackage{bigints}
\usepackage{dsfont}
\usepackage[dvipsnames]{xcolor}
\usepackage{epsfig}
\usepackage{epstopdf}
\usepackage{pdfpages}
\usepackage{pgffor}
\usepackage{graphicx}
\usepackage{multirow}
\usepackage{mathtools}
\usepackage[T1]{fontenc}
\usepackage{times} 
\usepackage{tensor}
\usepackage[utf8]{inputenc}
\usepackage[unicode=true,breaklinks=true,colorlinks=true]{hyperref}

\hypersetup{citecolor=blue,urlcolor=blue}

\newcommand\norm[1]{\left\lVert#1\right\rVert}

\makeatletter 
\AtBeginDocument{\let\LS@rot\@undefined} 
\makeatother 

\newcommand{\rev}[1]{\textcolor{black}{#1}}
\newcommand{\lxb}[1]{\textcolor{black}{#1}}

\usepackage[normalem]{ulem}

\newcommand{\dyad}[1]{\ket{#1}\bra{#1}}

\begin{document}

\title{Experimental demonstration of topological bounds in quantum metrology}

\author{Min Yu}
\thanks{These authors contributed equally.}
\author{Xiangbei Li}
\thanks{These authors contributed equally.}
\author{Yaoming Chu}
\email{yaomingchu@hust.edu.cn}
\affiliation{School of Physics, Hubei Key Laboratory of Gravitation and Quantum Physics, Institute for Quantum Science and Engineering, Huazhong University of Science and Technology, Wuhan 430074, China}
\affiliation{International Joint Laboratory on Quantum Sensing and Quantum Metrology, Huazhong University of Science and Technology, Wuhan 430074, China}
\author{Bruno Mera}
\email{bruno.mera.c5@tohoku.ac.jp}
\affiliation{Advanced Institute for Materials Research (WPI-AIMR), Tohoku University, Sendai 980-8577, Japan}
\author{F. Nur \"Unal}
\email{fnu20@cam.ac.uk}
\affiliation{TCM Group, Cavendish Laboratory, University of Cambridge, JJ Thomson Avenue, Cambridge CB3 0HE, United Kingdom\looseness=-1}
\author{Pengcheng Yang}
\affiliation{School of Physics, Hubei Key Laboratory of Gravitation and Quantum Physics, Institute for Quantum Science and Engineering, Huazhong University of Science and Technology, Wuhan 430074, China}
\affiliation{International Joint Laboratory on Quantum Sensing and Quantum Metrology, Huazhong University of Science and Technology, Wuhan 430074, China}
\author{Yu Liu}
\affiliation{Institut für Theoretische Physik and IQST, Albert-Einstein Allee 11, Universität Ulm, Ulm D-89081 Germany}
\affiliation{International Joint Laboratory on Quantum Sensing and Quantum Metrology, Huazhong University of Science and Technology, Wuhan 430074, China}
\author{Nathan Goldman}
\email{nathan.goldman@ulb.be}
\affiliation{Center for Nonlinear Phenomena and Complex Systems, Université Libre de Bruxelles, CP 231, Campus Plaine, B-1050 Brussels, Belgium}
\author{Jianming Cai}
\email{jianmingcai@hust.edu.cn}
\affiliation{School of Physics, Hubei Key Laboratory of Gravitation and Quantum Physics, Institute for Quantum Science and Engineering, Huazhong University of Science and Technology, Wuhan 430074, China}
\affiliation{International Joint Laboratory on Quantum Sensing and Quantum Metrology, Huazhong University of Science and Technology, Wuhan 430074, China}
\affiliation{Shanghai Key Laboratory of Magnetic Resonance, East China Normal University, Shanghai 200062, China}

\begin{abstract}
Quantum metrology is deeply connected to quantum geometry, through the fundamental notion of quantum Fisher information. Inspired by advances in topological matter, it was recently suggested that the Berry curvature and Chern numbers of band structures can dictate strict lower bounds on metrological properties, hence establishing a strong connection between topology and quantum metrology. In this work, we provide a first experimental verification of such topological bounds, by performing optimal quantum multi-parameter estimation and achieving the best possible measurement precision. By emulating the band structure of a Chern insulator, we experimentally determine the metrological potential across a topological phase transition, and demonstrate strong enhancement in the topologically non-trivial regime. Our work opens the door to metrological applications empowered by topology, with potential implications for quantum many-body systems.

\end{abstract}

\maketitle

\date{\today}

{\it Introduction.---} Exploring the limit of quantum precision measurement, as governed by the laws of quantum mechanics, not only leads to disruptive applications in quantum enhanced metrology \cite{Braunstein1994,Boixo2007,Demkowicz2015,Escher2011,Degen2017,Pezze2018,Giovannetti2011,Braun2018}, but also provides novel insights into fundamental concepts in quantum physics, such as entanglement, nonlocality and criticality  \cite{Sidhu2020,Zanardi2008,Hyllus2012,Hauke2016,Ma2009,Ma2009,Yadin2021,Niezgoda2021}. The precision limit for single-parameter estimation is given by the quantum Cram\'er-Rao bound (CRB) \cite{Braunstein1994}, which relates the best achievable measurement precision to the inverse of the quantum Fisher information (QFI) of the underlying quantum state. From a geometric perspective, the quantum Cramér-Rao bound set by the QFI for single-parameter-estimation is connected to the quantum metric~\cite{Braunstein1994,Sidhu2020}, which has recently been the focus of increased attention due to the recently established connection to flatband superconductivity
\cite{Peotta2015NC,Bernevig2022NRP}. This geometric property of quantum states corresponds to the real part of the quantum geometric tensor~\cite{Resta2011_EPJ}, which was also recently measured in experiments~\cite{asteria2019measuring,Yu2020NSR,Tan2019,Klees2020_PRL,Chen2022_Science}.  
More importantly, the imaginary part of the quantum geometric tensor corresponding to the Berry curvature plays a central role in topological physics, e.g. in quantum Hall-type transport ~\cite{Xiao2010,qi2011topological} and topological defects~\cite{armitage2018weyl}. Surprisingly, inspired by the existence of correlations between the quantum metric and the Berry curvature, it has been suggested that the Berry curvature (and the related Chern numbers) can set topological bounds on quantum multi-parameter estimation~\cite{meraRelatingTopologyDirac2022,Li2022}. Therefore, demonstrating the fundamental connection between topology and quantum metrology in experiments is highly appealing. While recent experiments realized and verified the CRB through QFI measurements \cite{Strobel-2014-Science,Pan2019,Rath_2021_PRL,Yu2022_PRR,Xu2022_PRL,Yu2022} in the context of single-parameter-estimation schemes \cite{Yu2022}, the extension to multi-parameter scenarios is generally more complex and challenging due to the possible incompatibility of optimal quantum measurements for each individual parameter \cite{Szczykulska2016,Albarelli2020,liuQuantumFisherInformation2020,Roccia2017,Polino2019,Hong2021,Hou2021b,Ciampini2016,Zhou2015,Ragy2016,Vidrighin2014}. Accessing the limits of quantum multi-parameter estimation has remained elusive, and the experimental demonstration of topological bounds in quantum metrology thus has never been explored. %

In this Letter, we address these challenges and present the first experiment connecting multi-parameter metrological bounds to topological band structures, using a synthetic topological system emulating a two-dimensional Chern insulator. By performing optimized positive operator-valued measurements (POVM) to implement quantum multi-parameter estimation of this synthetic topological system, we obtain the best achievable measurement precision. This allows us to experimentally verify the metrological bound given by the Berry curvature, and more importantly saturate the Holevo bound pertaining to geometric properties of the system. The developed techniques enable us to characterise quantum metrological potential across different topological regimes, which exhibits an appealing connection to the Chern number. Our results pave the way for considerations beyond the single-particle case, where the fundamental connection between quantum metrology, the Berry curvature and the Chern numbers of band structures, is anticipated to have an important impact in many-body settings with the precision of multi-parameter estimation dictated by the underlying topology.

\begin{figure}[t] 
\centering 
\includegraphics[width=8.6cm]{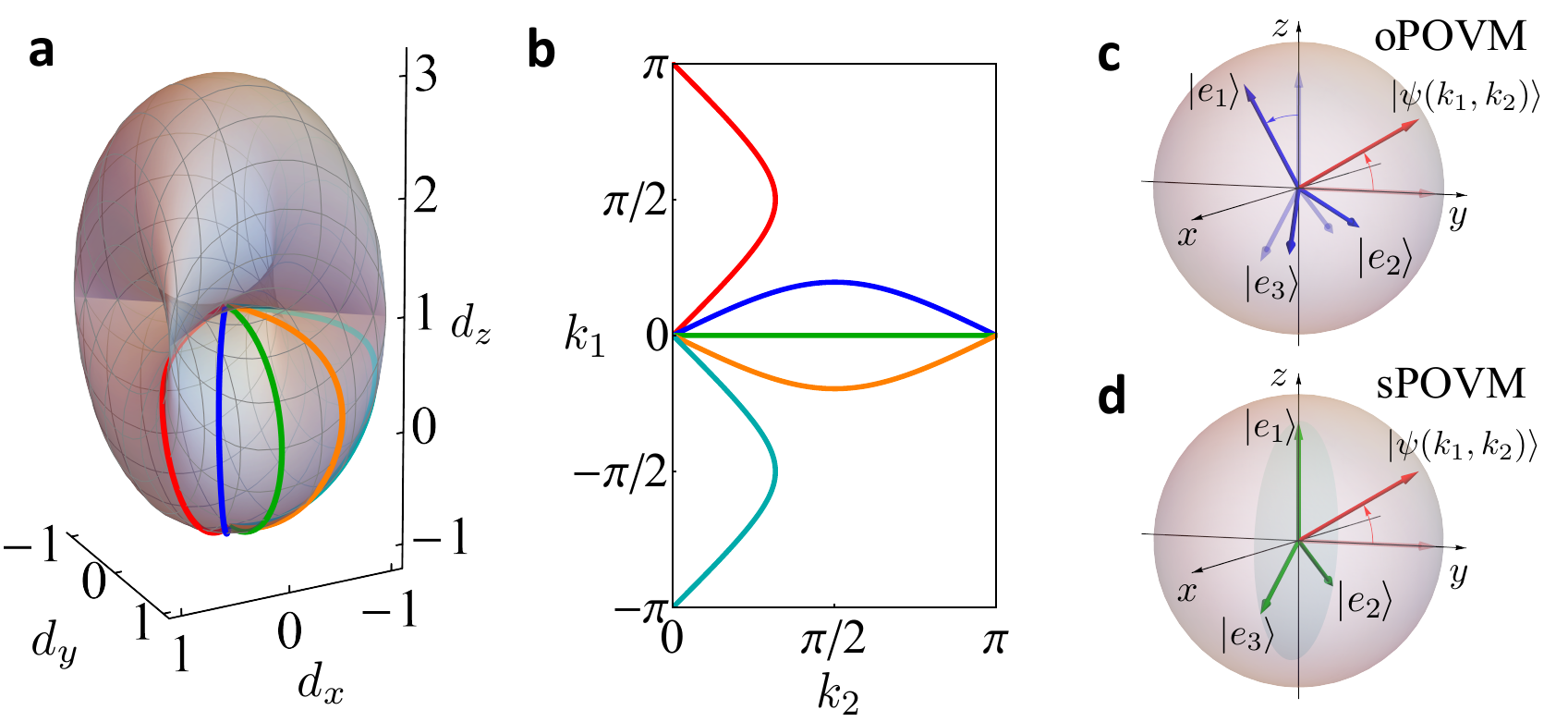} 
\caption{{Topology of a two-band Chern insulator and POVM for quantum multi-parameter estimation}. ({\bf a}) Surface of the terminal points $\boldsymbol{d}(\boldsymbol{k})$
[cf. Eq.(\ref{eq:chern-insulator-Hamiltonian}) with $M=1$]. ({\bf b}) The curves of different colors correspond to different trajectories in the Brillouin zone $\boldsymbol{k}=(k_1,k_2)\in \mathbb{T}^2$. ({\bf c}-{\bf d}) depict the optimized POVM (oPOVM) for different states ({\bf c}, blue arrows) and a fixed set of symmetric POVM ({\bf d}, sPOVM). The red arrows denote the Bloch vector of the excited state encoding the information of unknown parameters.}
\label{Topology} 
\end{figure}
{\it Quantum multi-parameter estimation of a synthetic topological system.---} General (and exact) relations between the quantum metric, the QFI and topological invariants exist for generic Dirac Hamiltonians in arbitrary spatial dimensions~\cite{meraRelatingTopologyDirac2022}. To experimentally investigate and verify these relations in detail, we utilize a nitrogen-vacancy (NV) center in diamond to implement a two-level synthetic topological system, which can describe a Chern insulator in two dimensions. The ground state of the NV center spin has three spin sublevels, $m_s=0,\pm 1$. By applying an external magnetic field along the NV axis, we lift the degeneracy of the spin states $m_s=\pm 1$ and employ the spin sublevels $m_s=0,-1$ to encode the two-level Hamiltonian; the additional spin state $m_s = +1$ is used for the implementation of POVM measurements \cite{supplement}. %
Our experiment aims at emulating the massive Dirac model~\cite{qi2006topological, qi2008topological}, given by the Bloch Hamiltonian
\begin{equation}
H(\boldsymbol{k})  =\boldsymbol{d}_{\boldsymbol{k}}\cdot\boldsymbol{\sigma} 
=\sum_{i=1}^2\sin(k_i)\sigma_i +\left(M-\sum_{i=1}^2\cos(k_i)\right)\sigma_3,
\label{eq:chern-insulator-Hamiltonian}
\end{equation}
where $\boldsymbol{d}_{\boldsymbol{k}}\in\mathbb{R}^3$ and $\boldsymbol{k}\in\mathbb{T}^2$ is the quasi-momentum. This model describes the band structure of a two-band Chern insulator [Fig.~\ref{Topology}(a-b)], exhibiting the quantum anomalous Hall effect. Away from the critical values of $M$ where the system is gapless, the vector $\boldsymbol{d}_{\boldsymbol{k}}$ gives rise to a well-defined unit vector $\boldsymbol{n}_{\boldsymbol{k}}=\boldsymbol{d}_{\boldsymbol{k}}/|\boldsymbol{d}_{\boldsymbol{k}}|\in \mathcal{S}^2$. 
The Hamiltonian in Eq.\eqref{eq:chern-insulator-Hamiltonian} is associated with two bands, with opposite Chern numbers and Berry curvature $\Omega_{12}(\boldsymbol{k})$. The aim of this work is to experimentally explore the connections between these geometric/topological quantities and multi-parameter estimation by performing the latter on a specific band. To achieve this goal, we estimate $\boldsymbol{k}$ by performing measurements on the eigenstates of a given band, in this case, the high energy band. We first prepare the system into the excited state $\vert \psi(\boldsymbol{k})\rangle$ of the Hamiltonian in Eq.~\eqref{eq:chern-insulator-Hamiltonian}, which encodes the unknown parameters $\boldsymbol{k}=(k_1,k_2)$. It is worth noting that two-outcome projective measurement is not sufficient to extract the complete information of both components $k_1$ and $k_2$ from the state $\vert \psi(\boldsymbol{k})\rangle$ \cite{supplement}. One needs to implement a generalized quantum measurement (namely a POVM), which can be specified by a set of operators as $\Pi=\{\Pi_i|\sum_i\Pi_i=\hat{\mathds{1}},\Pi_i\geq 0, i=1,\cdots,m\}$ on the system with $m \geq 3$ \cite{liuQuantumFisherInformation2020}. The results of $N$ measurement repetitions are represented as $\vec{x}=(x_{1},\,x_{2},\cdots,\,x_{k},\cdots,\,x_{N})$, where $x_k\in \{a_i\}_{i=1}^m$ and $a_i$ denotes the measurement outcome corresponding to $\Pi_i$. 
To optimally infer the vector $\boldsymbol{k}$, we construct the maximum likelihood estimator $\hat{\boldsymbol{k}}$ from the probability estimators $ \boldsymbol{\hat{p}}_\Pi(\vec{x})$, with $ \boldsymbol{\hat{p}}_\Pi^j(\vec{x}) =(1/N) \sum_{k=1}^N \delta_{a_j,x_{k}}$, by solving the likelihood equation \cite{supplement}. Consequently, the covariance matrix $\Sigma(\hat{\boldsymbol{k}})$ of the maximum likelihood estimator $\hat{\boldsymbol{k}}$ can be obtained as
\begin{align}
\Sigma(\hat{\boldsymbol{k}}) =\left( \frac{\partial\hat{\boldsymbol{k}}}{\partial \boldsymbol{\hat{p}}_\Pi} \right)  \Sigma(\boldsymbol{\hat{p}}_\Pi)\left( \frac{\partial\hat{\boldsymbol{k}}}{\partial \boldsymbol{\hat{p}}_\Pi} \right)^T,
    \label{eq-main:covariance}
\end{align}
where $\Sigma(\boldsymbol{\hat{p}}_\Pi)$ is the covariance matrix of $ \boldsymbol{\hat{p}}_\Pi$ and $\left(\frac{\partial\hat{\boldsymbol{k}}}{\partial \boldsymbol{\hat{p}}_\Pi}\right)$ is the associated Jacobian matrix. The square root of the generalized variance, $[\operatorname{det} \Sigma(\hat{\boldsymbol{k}})]^{1/2}$, measures the overall dispersion of multiple parameters, which we refer to as measurement uncertainty volume. This quantity is proportional to the volume of the hyper-elliptical estimated data cloud in $\hat{\boldsymbol{k}}$ space \cite{Tatsuoka1988}.
In our experiment, we adopt a set of 3-element rank-1 POVM $\{\Pi_i=\ket{e_i}\bra{e_i},~i=1,2,3\}$ that allows us to construct an unbiased estimator for two unknown parameters simultaneously (see \cite{supplement} for details). Such POVMs can be described with parameters $r_i$, $\theta_i$ and $\phi_i$ by setting 
\begin{equation}
\ket{e_i} = r_i \left( \cos{\frac{\theta_i}{2}} \ket{0} + \sin{\frac{\theta_i}{2}}e^{i\varphi_i} \ket{-1} \right).
\label{eq-main: ei}
\end{equation}
Note that the normalization condition of a POVM requires $\sum_{i=1}^3 r_i^2=2$, thus $\{\ket{e_i}\}_{i=1}^3$ is a set of unnormalized non-orthogonal vectors in the two-dimensional Hilbert space. This POVM is realized through a projective measurement $\{\dyad{u_i}\}_{i=1}^3$ in the extended three-level Hilbert space, where $(\dyad{0}+\dyad{-1})\ket{u_i} = \ket{e_i}$, by taking advantage of the auxiliary spin sublevel $m_s=+1$ of the NV center. To achieve this goal, we first apply unitary transformations on the NV center, which rotate the states $\ket{u_i}$ to the spin sublevels $\{\ket{0},\ket{\pm 1}\}$ by engineering microwave driving fields on resonance with both transitions $m_s=0 \leftrightarrow m_s=\pm 1$. The subsequent spin-dependent fluorescence measurement realizes projective measurements along the basis $\{\ket{u_i}\}$, which is equivalent to the  POVM $\{\Pi_i=\dyad{e_i}\}_{i=1}^3$ in the two-level Hilbert space spanned by $\{\ket{0},\ket{-1}\}$ \cite{supplement}. 
\begin{figure}[!t] 
\centering 
\includegraphics[width=8.6cm]{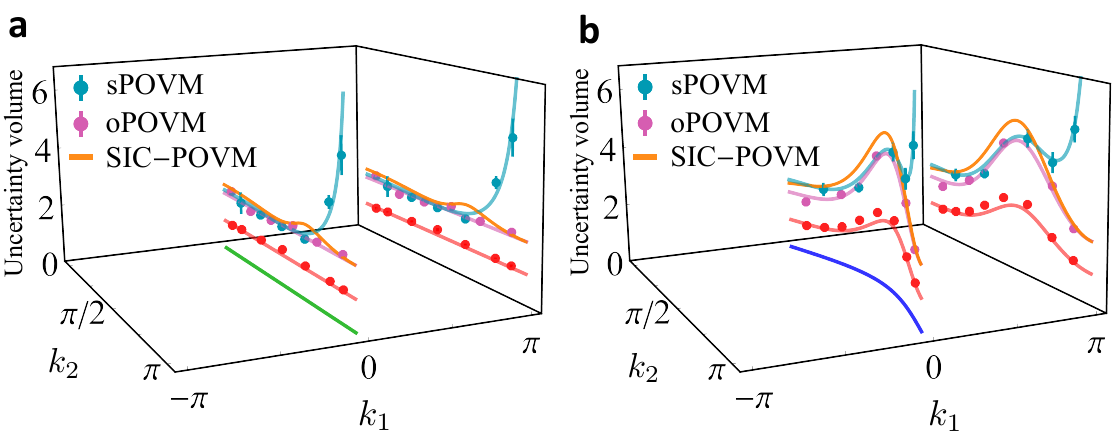} 
\caption{The measurement uncertainty volume $[\operatorname{det} \Sigma(\hat{\boldsymbol{k}})]^{1/2}$ as quantified by the square root of the generalized variance along two different trajectories in the $\boldsymbol{k}$-space, i.e. the green and blue curves \rev{[cf. Fig.~\ref{Topology}(a-b)]}.  The optimized POVM (oPOVM) achieves better performance over the state-independent symmetric POVM (sPOVM) and the symmetric, informationally complete POVM (SIC-POVM, theory), which are also compared with the bound given by the Berry curvature (red dots, the r.h.s. of Eq.\eqref{Eq:Sigma_Omega}).  The curves represent theoretical predictions. For better visibility, the curves and dots are projected to the side.}
\label{fig2} 
\end{figure}
The above appropriate parameterization of the POVM [i.e. Eq.~(\ref{eq-main: ei})] enables us to identify and implement a simple POVM that maximizes the determinant of the corresponding Fisher information matrix, which we denote as oPOVM \cite{supplement}; see Fig.~\ref{Topology}(c). In addition, we also implement a symmetric POVM (sPOVM) with 
$r_i=\sqrt{2/3}$, $\varphi_i=0$, and $\theta_i=0$, $\pm 2\pi/3$; see Fig.~\ref{Topology}(d). In Fig.~\ref{fig2}, we display the measurement uncertainty volume for the quantum multi-parameter estimation associated with the excited state $\ket{\psi(\boldsymbol{k})}$ of the synthetic topological Hamiltonian in Eq.~\eqref{eq:chern-insulator-Hamiltonian}. The experimental results are obtained by two different types of POVM measurements, namely the optimized POVM and the symmetric POVM. Here, we construct the probability estimator $\boldsymbol{\hat{p}}_\Pi(\vec{x})$ from the experimental measurement outcomes, and obtain the covariance matrix $\Sigma(\hat{\boldsymbol{k}})$ according to Eq.~\eqref{eq-main:covariance} \cite{supplement}. For comparison, we also present the values that can be achieved via the symmetric, informationally complete POVM (SIC-POVM) \cite{Renes2004,LiNan2016,axioms6030021,ZhuHuangjun2018}, which represents the most versatile class of measurements to obtain information about the state of a quantum system. It can be seen from Fig.~(\ref{fig2}) that the optimized POVM that we identify indeed achieves better measurement performance over both symmetric POVM and SIC-POVM. 
{\it Optimal quantum multi-parameter estimation and topological bounds.---} The above developed techniques enable us to experimentally explore the metrological bounds related to the topology of the system. We note that the multi-parameter CRB establishes a lower bound for the covariance matrix \cite{liuQuantumFisherInformation2020}
\begin{equation}
\Sigma(\hat{\boldsymbol{k}})\geq  \frac{1}{N} \mathcal{F}^{-1}_{\mathbb{T}^2},
\end{equation}
where \lxb{$N$ represents the number of measurements and} $\mathcal{F}_{\mathbb{T}^2}$ is the QFI matrix of $\ket{\psi(\boldsymbol{k})}$ with respect to the vector $\boldsymbol{k}$. Remarkably, the Berry curvature $\Omega_{12}(\boldsymbol{k})$ associated with the state $\ket{\psi(\boldsymbol{k})}$ is related to the quantum metric (and thereby the QFI matrix as $\mathcal{F}_{\mathbb{T}^2} =4 g(\boldsymbol{k})$) through $[{\mathrm{det}(g(\boldsymbol{k})})]^{1/2}=\left|\Omega_{12}(\boldsymbol{k})\right|/2$. This surprisingly concise identification has important metrological implications: the uncertainty volume for quantum parameter estimation is bounded by the Berry curvature as~\cite{meraRelatingTopologyDirac2022}
\begin{equation}
\label{Eq:Sigma_Omega}
[\operatorname{det} \Sigma(\hat{\boldsymbol{k}})]^{1/2} >\frac{1}{2N}\frac{1}{\left|\Omega_{12}(\boldsymbol{k})\right|}.
\end{equation}
In our experiment, we obtain the measurement uncertainty volume $[\operatorname{det} \Sigma(\hat{\boldsymbol{k}})]^{1/2}$ achieved by the optimized POVM, and extract the full quantum geometric tensor (QGT) by measuring the fidelity between neighbouring quantum states in parameter space \cite{supplement,Yu2022_PRR}. This allows to directly compare the best achievable measurement uncertainty volume with the Berry curvature bound [namely the r.h.s in Eq.~\eqref{Eq:Sigma_Omega}]. The results shown in Fig.~\ref{fig2}(a-b) not only experimentally verify the Berry curvature bound [Eq.(\ref{Eq:Sigma_Omega})] for the first time, but also suggest that the achieved optimal measurement uncertainty volume shows strongly correlated behavior with the Berry curvature bound. This implies that a larger Berry curvature is associated with a better metrological performance (namely a smaller measurement uncertainty volume). Hence, our experiment demonstrates how extracting the Berry curvature  -- an effective magnetic field in momentum space~\cite{Xiao2010} -- provides a practical scheme to predict the metrological potential of topological band systems.
\begin{figure}[!t] 
\centering 
\includegraphics[width=8.6cm]{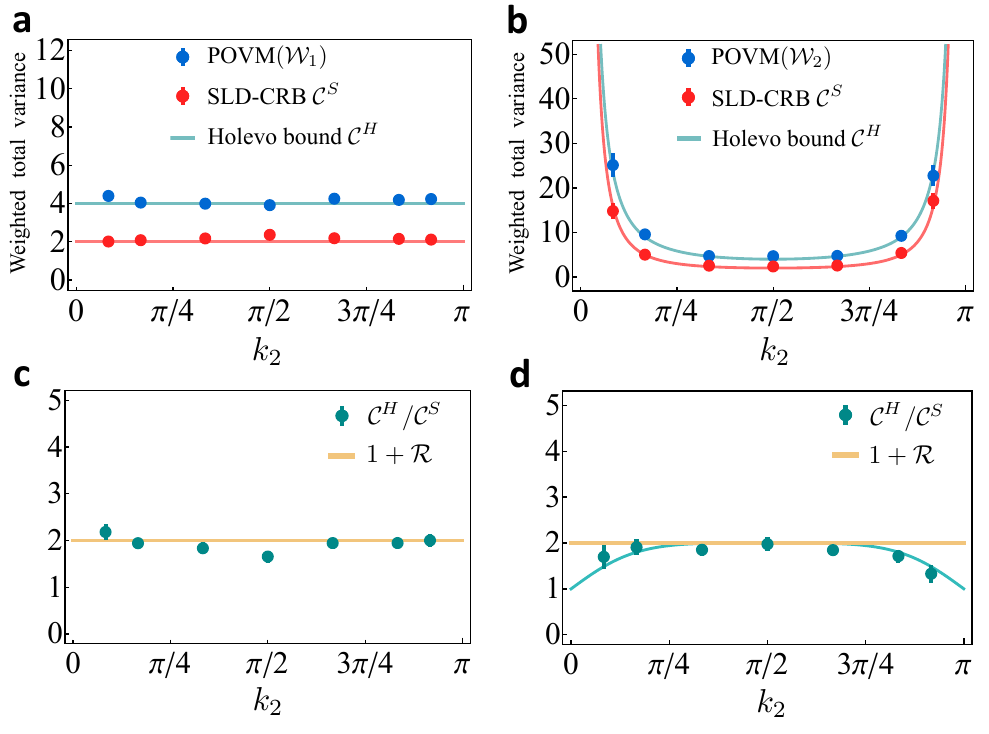}
\caption{Scalar CRBs for quantum multi-parameter estimation, {along a trajectory in the $\boldsymbol{k}$-space, i.e. the green curve in Fig.\ref{Topology}(a-b),} with respect to different weight matrices $\mathcal{W}_1=\mathcal{F}_{\mathbb{T}^2}$ ({\bf a}) and $\mathcal{W}_2=J^T J$ ({\bf b}). The POVM($\mathcal{W}_j$) ($j=1,2$), which is optimized to achieve the minimal value of $\operatorname{Tr}(\mathcal{W}_j F_C^{-1})$ for the Fisher information matrix $F_C$, saturates the Holevo bound, $\mathcal{C}^H(\boldsymbol{k},\mathcal{W}_j)$. ({\bf c-d}) show the ratio between the Holevo bound and the SLD-CRB $(\mathcal{C}^H/\mathcal{C}^S)$ for the weight matrices $\mathcal{W}_1$ ({\bf c}) and $\mathcal{W}_2$ ({\bf d}), which is compared with the characterization parameter $1+\mathcal{R}$. The curves represent theoretical predictions.} 
\label{fig3} 
\end{figure}
In addition to the measurement uncertainty volume, the precision for multi-parameter estimation is characterized by the weighted total variance $\operatorname{Tr}(\mathcal{W}\Sigma(\hat{\boldsymbol{k}})$ with a positive real weight matrix $\mathcal{W}$. The achievable measurement precision limit as quantified by the weighted total variance is given by the Holevo bound (referred to as the attainable quantum CRB) \cite{demkowicz-dobrzanski2020,supplement}, which can only be obtained as an optimization. The techniques that we develop for the optimization and implementation of POVM allow us to achieve such a non-trivial goal. In the experiment, for a chosen weight matrix $\mathcal{W}_j$, we perform the optimized POVM, the corresponding Fisher information matrix ($F_C$) of which minimizes the value of $\operatorname{Tr}(\mathcal{W}_j F_C^{-1})$, and obtain the covariance matrix $\Sigma(\hat{\boldsymbol{k}})$. As demonstrated in Fig.~\ref{fig3}(a), when we choose the weight matrix $\mathcal{W}_1=\mathcal{F}_{\mathbb{T}^2}$, the achieved scalar measurement uncertainty $\operatorname{Tr}(\mathcal{W}_1 \Sigma(\hat{\boldsymbol{k}}))$ indeed reaches the corresponding Holevo bound $\mathcal{C}^H(\boldsymbol{k},\mathcal{W}_1)$. As a second example, we consider the weight matrix $\mathcal{W}_2=J^{T}J$, where $J$ is the Jacobian matrix associated with the pullback map from the 2-sphere $\mathcal{S}^2$ to the Brillouin zone $\mathbb{T}^2$\cite{meraRelatingTopologyDirac2022,supplement}. Similarly, we perform the optimized POVM with respect to the weight matrix $\mathcal{W}_2$, which also saturates the corresponding Holevo bound [Fig.~\ref{fig3}(b)].

Remarkably, the Holevo bound has significant geometric relevance \cite{Matsumoto2000,Matsumoto2002,Li2022}, and is connected (via the Berry curvature) with the quantum SLD-CRB $\mathcal{C}^S(\boldsymbol{k},\mathcal{W})\equiv \operatorname{Tr}(\mathcal{W}\mathcal{F}_{\mathbb{T}^2}^{-1})$, namely $\mathcal{C}^H(\boldsymbol{k},\mathcal{W})\leq (1+\mathcal{R})\mathcal{C}^S(\boldsymbol{k},\mathcal{W})$~\cite{carollo:19,Albarelli2020,Li2022}. The parameter $\mathcal{R}=\norm{i2\mathcal{F}_{\mathbb{T}^2}^{-1}\Omega}_{\infty}\in [0,1]$ is related with the Berry curvature $\Omega$, with $\norm{\bullet}_{\infty}$ taking the largest eigenvalue \cite{Li2022}. We determine the quantum SLD-CRB $\mathcal{C}^S(\boldsymbol{k},\mathcal{W})$ for the weighted total variance by measuring quantum metric $\sim \mathcal{F}_{\mathbb{T}^2}/4$ \cite{supplement,Yu2022_PRR}. The results in Fig.~\ref{fig3}(c-d) show the ratio between the Holevo bound and the quantum SLD-CRB $\mathcal{C}^{H}(\boldsymbol{k},\mathcal{W})/\mathcal{C}^S(\boldsymbol{k},\mathcal{W})$ and directly testify the attainability of the quantum SLD-CRB.

\begin{figure}[!t] 
\centering 
\includegraphics[width=8.8cm]{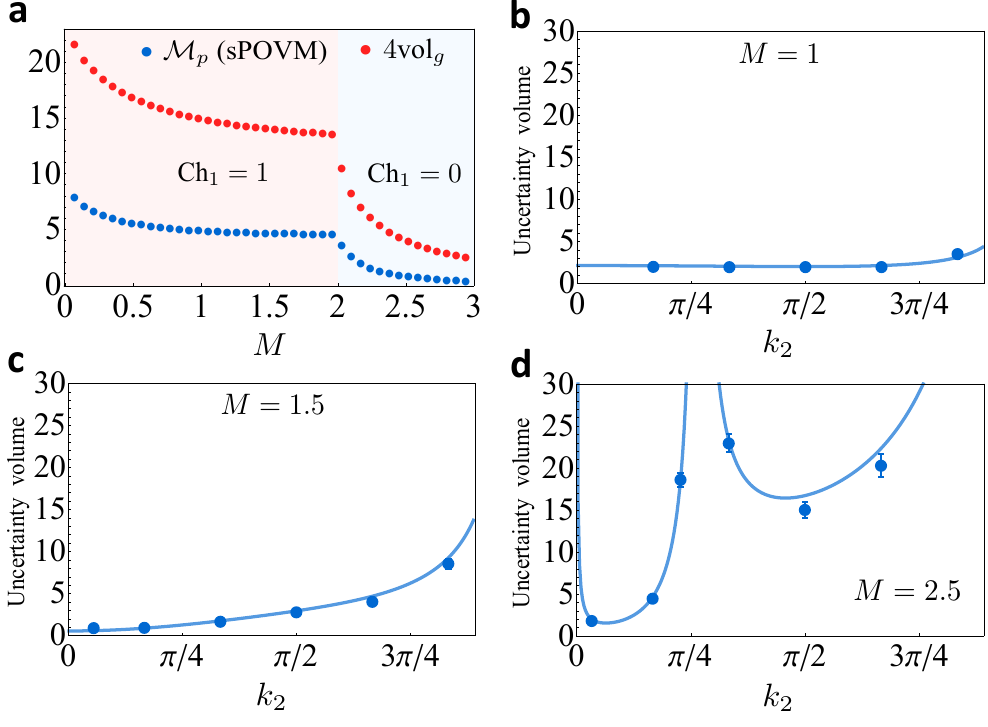} 
\caption{({\bf a}) Quantum volume $4\mathrm{vol}_g (\mathbb{T}^2)$ is compared with the integration of the inverse of the uncertainty volume $\mathcal{M}_p\equiv (1/N)\int_{\mathbb{T}^{2 }} [\operatorname{det} \Sigma(\hat{\boldsymbol{k}}) ]^{-1/2} \mathrm{d}  \boldsymbol{k}$ by sPOVM. ({\bf b}-{\bf d}) show the measurement uncertainty volume $[\operatorname{det} \Sigma(\hat{\boldsymbol{k}}) ]^{1/2}$ obtained by sPOVM along a trajectory in the $\boldsymbol{k}$-space [i.e. the green curve in Fig.\ref{Topology}(a-b)] for topologically different regimes. The blue dots represent experimental data, while the theoretical values are indicated with lines.}
\label{fig4} 
\end{figure}

We remark that the Berry curvature bound and the attainability of the quantum SLD-CRB (as we metrologically characterize in experiments) reveal the intriguing role of the Berry curvature in determining metrological potential of topological systems: A larger Berry curvature would be beneficial for the measurement precision of quantum multi-parameter estimation, however it may indicate a weaker attainability of the quantum SLD-CRB.
{\it Metrological characterization of topological bands.---} Furthermore, we experimentally explore the metrological potential of the Bloch Hamiltonian [Eq.\eqref{eq:chern-insulator-Hamiltonian}] in different topological regimes governed by the control parameter $M$, where $|M|>2$ and $|M|<2$ correspond to topologically trivial (with the Chern number $\mathrm{Ch}_{1}=0$) and non-trivial ($\mathrm{Ch}_{1}=1$) cases, respectively. The quantum volume of the momentum space, which is sensitive to the topology, is defined as $\mathrm{vol}_g (\mathbb{T}^2)\equiv \int_{\mathbb{T}^{2 }} [\mathrm{det}(g(\boldsymbol{k}))]^{1/2}\mathrm{d} \boldsymbol{k}$ \cite{meraRelatingTopologyDirac2022}. In Fig.\ref{fig4}(a), we display the integrated measurement uncertainty using the symmetric POVM over the Brillouin zone, which shows correlated behaviour with the quantum volume across the topological transition at $M=2$. The result confirms that the quantum volume can predict the metrological potential of a topological system following \cite{supplement}
\begin{equation}
\mathcal{M}_p \equiv (1/N)\int_{\mathbb{T}^{2 }} [\operatorname{det} 
\Sigma(\hat{\boldsymbol{k}}) ]^{-1/2} \mathrm{d}k_1 \mathrm{d}k_2\leqslant 4 \mathrm{vol}_g (\mathbb{T}^2),
\end{equation}
namely the larger quantum volume $\mathrm{vol}_g (\mathbb{T}^2)$ implies the better overall metrological performance in the Brillouin zone.
The integration of the metric--Berry curvature relation over the Brillouin zone links the quantum volume $\mathrm{vol}_g (\mathbb{T}^2)$ to a topological invariant, namely the first Chern number $\mathrm{Ch}_{1}$ of the related Bloch band, via $\mathrm{vol}_g(\mathbb{T}^2)\geq \pi |\mathcal{C}|$ \cite{meraRelatingTopologyDirac2022}. The equality holds if the Berry curvature keeps the same sign over the Brillouin zone \cite{meraRelatingTopologyDirac2022}. This relation further predicts that the system's metrological performance may strongly depend on its topological invariants.
We proceed to choose $M=1$ and $M=1.5$ in the topologically non-trivial regime and $M=2.5$ in the topologically trivial regime, and experimentally determine the measurement uncertainty volume to illustrate the corresponding metrological performance; see Fig.~\ref{fig4}(b-d). These results clearly demonstrate the contrast in metrological potentials of the topologically different regimes: the measurement uncertainty volume is significantly smaller in the topologically non-trivial regime ($\mathrm{Ch}_{1}=1$) than in the topologically trivial regime ($\mathrm{Ch}_{1}=0$). Our experiment, hence, provides clear evidence that topology influences metrological potential in a non-trivial way.
{\it Conclusion \& outlook.---} We have demonstrated quantum multi-parameter estimation in a synthetic topological system realized by a highly-controllable NV center spin in diamond. By optimizing and implementing POVMs to extract information on two parameters simultaneously, we have achieved the best possible measurement precision characterized by the uncertainty volume and the weighted total variance. We have thus verified the metrological bound set by the Berry curvature, and saturated the Holevo bound (namely the attainable quantum CRB) \rev{accessing the limits} of quantum multi-parameter estimation. Furthermore, we have experimentally explored the enhanced metrological potential across topological phase transitions. Our work establishes a fundamental connection between quantum metrology and the geometric features of topological band structures. 
The developed methods can be extended to a variety of topological systems, including many-body quantum systems. As a gedankenexperiment, which could be realized in quantum-engineered systems~\cite{manovitz2020quantum}, one considers a two-band Chern insulator on a torus geometry, with non-interacting fermions filling the lowest band. By threading magnetic fluxes $\phi_{1,2}$ through two non-contractible cycles of the torus, one obtains a family of many-body ground states over $\phi_{1,2}$-space \cite{Niu1985_PRB,Kudo2019_PRL}. These magnetic fluxes can be estimated by performing measurements on the many-body wave function, which represents a metrological task of multi-parameter quantum estimation. According to our results, this metrological task should be correlated to the (many-body) Chern number of the Chern insulator. Other connections between  metrological performances and the topology of many-body systems can be established through edge properties~\cite{Sarkar2022}.

{\it Note added.---} During the preparation of this manuscript, the authors became aware of the related work~\cite{Li2022}, which experimentally investigates the quantum geometry of quantum multi-parameter sensing.

{\it Acknowledgments.---} We thank Prof. Lijiang Zhang for helpful discussions on quantum multi-parameter estimation. The work is supported by the National Natural Science Foundation of China (Grant No.~12161141011, 11874024), the National Key R$\&$D Program of China (Grant No. 2018YFA0306600), Shanghai Key Laboratory of Magnetic Resonance (East China Normal University), the FRS-FNRS (Belgium), the ERC (Starting Grant TopoCold), the Royal Society under a Newton International Fellowship, the Marie Sk{\l}odowska-Curie programme of the European Commission Grant No 893915,  the EOS (CHEQS project). Y. L was supported by the BMBF under the funding program quantum technologies-from basic research to market in the project Spinning (Grant No. 13N16215).

%

%

\bibliography{reference}




\section*{Supplementary material}

# Supplementary Materials

## Bounds for quantum multi-parameter estimation

The quantum Cramér-Rao bound (QCRB) sets a fundamental limit on the accuracy of unbiased quantum parameter estimation, relating the uncertainty in determining parameters to the inverse of the quantum Fisher information (QFI). The QCRB is generally achievable in the case of single parameter estimation (see e.g. Ref. [1]). However, its multi-parameter version is not always attainable due to the incompatibility between optimal measurement operators for different parameters, where the QFI is generalized to QFI matrix (QFIM) [2].

### Quantum SLD Cramér-Rao bound

The QCRB states that the measurement uncertainty of the parameters $\boldsymbol{\lambda} = (\lambda_1, \cdots, \lambda_d)$ in dimension $d$, as characterized by the covariance matrix of the unbiased estimator [2] $\Sigma(\hat{\boldsymbol{\lambda}}) = [\langle \delta\hat{\lambda}_i \delta\hat{\lambda}_j \rangle]_{1 \leq i,j \leq d}$ is bounded by the QFIM, $\mathcal{F}_Q(\boldsymbol{\lambda})$, which is related to the quantum metric, $g(\boldsymbol{\lambda}) = [g_{ij}(\boldsymbol{\lambda})]_{1 \leq i,j \leq d}$, through the inequality

$$\Sigma(\hat{\boldsymbol{\lambda}}) \geqslant \frac{1}{N}\mathcal{F}_Q^{-1}(\boldsymbol{\lambda}) = \frac{1}{4N}g^{-1}(\boldsymbol{\lambda}), \tag{S.1}$$

where $N$ is the number of measurement repetitions and note that the last equal sign of the above equation holds for the pure state. In the following, we will refer to this bound as the quantum SLD-CRB, due to the fact that the QFI matrix is defined via the symmetric logarithmic derivative (SLD) operator. To be precise, the symmetric logarithmic derivative of a smooth family of density matrices $\rho_{\boldsymbol{\lambda}}$ is the matrix valued one-form, $L = \sum_i L_i d\lambda_i$ defined by

$$d\rho = \frac{1}{2}\left(L\rho + \rho L\right), \tag{S.2}$$

where $d = \sum_i d\lambda_i \frac{\partial}{\partial \lambda_i}$ is the exterior derivative. The quantum Fisher metric is then given by

$$[\mathcal{F}_Q(\boldsymbol{\lambda})]_{ij} = \frac{1}{2}\operatorname{Tr}\left(\rho_{\boldsymbol{\lambda}}\{L_i, L_j\}\right), \ i,j \in \{1, \ldots, d\}, \tag{S.3}$$

where $\{\cdot, \cdot\}$ is the matrix anti-commutator.

In order to quantify the overall performance of quantum multi-parameter estimation, it is convenient to recast the covariance matrix $\Sigma(\hat{\boldsymbol{\lambda}})$ into scalar characteristic functions. One representative scalar characteristic function is given by the generalized variance, i.e. $\det(\Sigma(\hat{\boldsymbol{\lambda}}))$. Based on Eq. (S.1), the bound for the generalized variance is given by

$$\sqrt{\det\left(\Sigma(\hat{\boldsymbol{\lambda}})\right)} \geqslant N^{-d/2}\sqrt{\det\left(\mathcal{F}_Q^{-1}(\boldsymbol{\lambda})\right)} = (4N)^{-d/2}\frac{1}{\sqrt{\det(g(\boldsymbol{\lambda}))}}. \tag{S.4}$$

Another typical scalar characteristic function corresponds to the weighted total variance, which is bounded by

$$\operatorname{Tr}\left(\mathcal{W}\Sigma(\hat{\boldsymbol{\lambda}})\right) \geqslant \frac{1}{N}\mathcal{C}^S(\boldsymbol{\lambda}, \mathcal{W}) \equiv \frac{1}{N}\operatorname{Tr}\left(\mathcal{W}\mathcal{F}_Q^{-1}(\boldsymbol{\lambda})\right) = \frac{1}{4N}\operatorname{Tr}\left(\mathcal{W}g^{-1}(\boldsymbol{\lambda})\right), \tag{S.5}$$

where $\mathcal{W}$ is a real symmetric matrix. We remark that the attainability of the quantum SLD-CRB is determined by whether it is possible to find a set of POVM $\{M_m\}$ for which the corresponding probabilities $p_m = \operatorname{Tr}(\rho_{\boldsymbol{\lambda}} M_m)$ yield the Fisher information matrix $F_C = \mathcal{F}_Q$, where $[F_C(\boldsymbol{\lambda})]_{ij} \equiv \sum_m p_m(\boldsymbol{\lambda})[\partial \ln p_m(\boldsymbol{\lambda})/\partial \lambda_i][\partial \ln p_m(\boldsymbol{\lambda})/\partial \lambda_j]$. In the single parameter case, such a set of POVM can always be constructed based on the SLD operator. However, the question is much more complex for the quantum multi-parameter estimation. If all the SLDs corresponding to different parameters commute with each other, one may saturate the quantum SLD-CRB by performing a joint measurement of the SLDs. But if the SLDs do not commute, POVMs that are optimal for different parameters are usually incompatible. This observation has led to the increasingly intensive investigation of more effective bounds for quantum multi-parameter estimation.

*Berry-curvature bound for generic Dirac systems*

Quantum geometry has emerged as a central and ubiquitous concept in quantum sciences, with direct consequences on quantum metrology and many-body quantum physics. The Fubini-Study metric introduces a notion of distance between quantum states defined over the parameter space, and the Berry curvature plays a crucial role in capturing Berry phase effects and topological band structures. The work in Ref. [3] establishes a general and exact relation between these two representative geometric quantities for generic Dirac Hamiltonians as,

$$\frac{i^n}{(2\pi)^n n!} \frac{1}{2^n} \sum_{i_1,j_1,\ldots,i_n,j_n=1}^{2n} \text{Tr}\left(\Omega_{i_1 j_1} \ldots \Omega_{i_n j_n}\right) \varepsilon^{i_1 j_1 \ldots i_n j_n} = \text{sgn}(d\vec{n})(-1)^n \frac{(2n)!}{2^{n(n-1)+1} n! \pi^n} \sqrt{\det(g)}, \tag{S.6}$$

where $\Omega$ is the Berry curvature, $n = \lfloor D/2 \rfloor$ and $D$ is the number of generators of the complex Clifford algebra. This relationship establishes a general topological bound for the generalized variance in Eq.(S.4),

$$\sqrt{\det(\Sigma(\hat{\boldsymbol{k}}))} > \frac{(2n)!}{N^n 2^{n^2-n+1}} \frac{1}{\left|\sum_{i_1,j_1,\ldots,i_n,j_n=1}^{2n} \text{Tr}\left(\Omega_{i_1 j_1} \ldots \Omega_{i_n j_n}\right) \varepsilon^{i_1 j_1 \ldots i_n j_n}\right|}, \tag{S.7}$$

where the momentum parameter $\boldsymbol{k}$ is defined over a $d$-dimensional Brillouin zone, $\mathbb{T}^d$, with $d = 2n$. As an example, we consider the two-dimensional case where the above inequality reduces to [i.e., Eq.(6) in the main text]

$$\sqrt{\det(\Sigma(\hat{\boldsymbol{k}}))} > \frac{1}{2N} \frac{1}{|\Omega_{12}(\boldsymbol{k})|}. \tag{S.8}$$

Such a topological bound given by the Berry curvature for quantum multi-parameter estimation can be interpreted as a constraint on the momentum estimation by an effective magnetic field in $\boldsymbol{k}$-space. In the main text (see Fig.2), we verify this Berry curvature bound in Eq.(S.8) by experimentally determining the generalized variance $\det(\Sigma(\hat{\boldsymbol{k}}))$ and the quantum geometric tensor (i.e. the quantum metric $g$ and the Berry curvature $\Omega$) separately.

*Holevo Cramér-Rao bound*

The Holevo Cramér-Rao bound (HCRB) is the asymptotically tight bound for general multi-copy estimation models. Given any measurement $M_m$ and an unbiased estimator $\hat{\boldsymbol{\lambda}}$ for the parameter $\boldsymbol{\lambda}$, we define a vector of Hermitian matrices $\mathbf{X} = [X_1, \ldots, X_d]^{\text{T}}$ as [4]

$$\mathbf{X} := \sum_m \left(\hat{\boldsymbol{\lambda}}(m) - \boldsymbol{\lambda}\right) M_m, \tag{S.9}$$

which satisfy the following conditions

$$\text{Tr}\left(\rho_{\boldsymbol{\lambda}} \mathbf{X}\right) = 0, \quad \text{Tr}\left(\boldsymbol{\nabla} \rho_{\boldsymbol{\lambda}} \mathbf{X}^{\text{T}}\right) = \mathcal{I}. \tag{S.10}$$

The Holevo bound is defined by [4]

$$\text{Tr}(\mathcal{W}\Sigma(\boldsymbol{\lambda})) \geqslant \frac{1}{N}\mathcal{C}^H(\boldsymbol{\lambda}, \mathcal{W}) \equiv \frac{1}{N} \min_{\mathbf{X}, V} \left(\text{tr}(\mathcal{W}V) \mid V \geqslant Z[\mathbf{X}], \quad \text{Tr}\left(\nabla \rho_{\boldsymbol{\lambda}} \mathbf{X}^{\text{T}}\right) = \mathcal{I}\right) \tag{S.11}$$

where $V$ is a $d \times d$ real matrix and $Z[\mathbf{X}] = \text{Tr}\left(\rho_{\boldsymbol{\lambda}} \mathbf{X} \mathbf{X}^{\text{T}}\right)$ is a $d \times d$ complex matrix. In our experiments, we saturate the Holevo bound by performing optimal POVM that minimizes the scalar function $\text{Tr}(\mathcal{W}F_C^{-1})$, see Fig.3 in the main text.

The attainable multi-parameter HCRB and the SLD-CRB satisfy the following relationship [5]

$$\begin{aligned}
\mathcal{C}^S(\boldsymbol{\lambda}, \mathcal{W}) &\leq \mathcal{C}^H(\boldsymbol{\lambda}, \mathcal{W}) \\
&\leq \mathcal{C}^S(\boldsymbol{\lambda}, \mathcal{W}) + \left\|\sqrt{\mathcal{W}} \boldsymbol{\mathcal{F}}_Q^{-1} \boldsymbol{D} \boldsymbol{\mathcal{F}}_Q^{-1} \sqrt{\mathcal{W}}\right\|_1 \\
&\leq (1 + \mathcal{R})\mathcal{C}^S(\boldsymbol{\lambda}, \mathcal{W}) \\
&\leq 2\mathcal{C}^S(\boldsymbol{\lambda}, \mathcal{W})
\end{aligned} \tag{S.12}$$

where $\|A\|_1 = \text{Tr}[\sqrt{A^\dagger A}]$ and $\boldsymbol{D}$ is the (asymptotic) incompatibility matrix (i.e. mean Uhlmann curvature) with element $D_{ij} = -i\,\text{Tr}\,[\rho_\lambda\,[L_i, L_j]]\,/4$ (for pure states this reduces to the Berry curvature $\Omega$). The parameter $\mathcal{R} = \left\|i2\mathcal{F}_Q^{-1}\Omega\right\|_\infty \in [0,1]$ characterizes the attainability of the scalar SLD-CRB [5–7], where $\|A\|_\infty$ denotes the largest eigenvalue of $A$. The ratio $\mathcal{C}^H/\mathcal{C}^S$ quantifies the discrepancy between the HCRB and the SLD-CRB which can be written as

$$1 \leq \mathcal{C}^H/\mathcal{C}^S \leq 1 + \mathcal{R} \leq 2. \tag{S.13}$$

Our experiment provides the lower bound of $\mathcal{R}$ by direct metrological determination of $\mathcal{C}^H(\boldsymbol{k},\mathcal{W})$ and geometric measurement of $\mathcal{C}^S(\boldsymbol{k},\mathcal{W})$, see Fig.3 (c-d) in the main text.

**Parameterization and realization of POVM**

*Parameterization of POVM*

Given a measurement observable, one can use the maximum likelihood estimator to extract the information on the unknown parameters. The corresponding covariance matrix $\Sigma$ is given by the inverse of the Fisher information matrix, i.e. $F_C^{-1}$. The performance of quantum parameter estimation is generally dependent on the measurement observable. In order to achieve the best metrological performance, we proceed to find the optimal POVM by minimizing the values of $\sqrt{\det F_C^{-1}}$ and $\text{Tr}(\mathcal{W}F_C^{-1})$.

We remark that in order to estimate $n$ parameters, the POVM usually needs to have at least $(n+1)$ elements, e.g. 3-element POVM may allow us to estimate two parameters simultaneously. Moreover, we find that the Holevo bound actually can be attained by optimizing the 3-element rank-1 POVM. Therefore, we start by considering a set of 3-element rank-1 POVM,

$$\Pi_i = |e_i\rangle\langle e_i|, \quad i = 1, 2, 3, \tag{S.14}$$

which satisfies the following normalization condition

$$\sum_{i=1}^{3} \Pi_i = \mathbb{1}. \tag{S.15}$$

The state vector $|e_i\rangle$ can be parameterized as

$$|e_i\rangle = r_i \left(\cos\frac{\theta_i}{2}|0\rangle + \sin\frac{\theta_i}{2}e^{i\varphi_i}|-1\rangle\right), \tag{S.16}$$

where $|0\rangle$ and $|-1\rangle$ denote the spin sublevel $m_s = 0$ and $m_s = -1$ of the NV center respectively. Based on Eq. (S.16), nine parameters are needed to determine a set of 3-element rank-1 POVM. The normalization condition in Eq. (S.15) leads to

$$\begin{aligned}
&r_1^2 + r_2^2 + r_3^2 = 2, \\
&r_1^2 \cos\theta_1 + r_2^2 \cos\theta_2 + r_3^2 \cos\theta_3 = 0, \\
&r_1^2 \sin\theta_1 \cos\varphi_1 + r_2^2 \sin\theta_2 \cos\varphi_2 + r_3^2 \sin\theta_3 \cos\varphi_3 = 0, \\
&r_1^2 \sin\theta_1 \sin\varphi_1 + r_2^2 \sin\theta_2 \sin\varphi_2 + r_3^2 \sin\theta_3 \sin\varphi_3 = 0.
\end{aligned} \tag{S.17}$$

which further reduces the number of free parameters in Eq. (S.16) to five. In addition, the last three equations of Eq. (S.16) tells that the three vectors defined by $\vec{n}_i = (\sin\theta_i \cos\varphi_i, \sin\theta_i \sin\varphi_i, \cos\theta_i)$ $(i=1,2,3)$ are in the same plane, i.e. $r_1^2 \vec{n}_1 + r_2^2 \vec{n}_2 + r_3^2 \vec{n}_3 = 0$. Next, we first introduce the following three state vectors in the $x-y$ plane,

$$\begin{aligned}
|e_{10}\rangle &= \frac{r_1}{\sqrt{2}}\left(|0\rangle + |-1\rangle\right), \\
|e_{20}\rangle &= \frac{r_2}{\sqrt{2}}\left(|0\rangle + e^{-i\vartheta}|-1\rangle\right), \\
|e_{30}\rangle &= \frac{r_3}{\sqrt{2}}\left(|0\rangle + e^{i\phi}|-1\rangle\right),
\end{aligned} \tag{S.18}$$

with

$$\begin{aligned}
\vartheta &= \arccos\left[(r_3^4 - r_1^4 - r_2^4)/(2r_1^2 r_2^2)\right], \tag{S.19} \\
\phi &= \arccos\left[(r_2^4 - r_1^4 - r_3^4)/(2r_1^2 r_3^2)\right]. \tag{S.20}
\end{aligned}$$

Only two parameters of $\{r_1, r_2, r_3\}$ are independent, without loss of generality, we set $r_1 \leqslant r_2 \leqslant r_3$ and $r_3 = (2 - r_1^2 - r_2^2)^{1/2}$, which satisfy the following conditions

$$0 \leqslant r_1 \leqslant \sqrt{\frac{2}{3}},$$
$$\max\left(r_1, \sqrt{1 - r_1^2}\right) \leqslant r_2 \leqslant \sqrt{1 - \frac{r_1^2}{2}}. \tag{S.21}$$

The general state vectors $|e_i\rangle$ in Eq.(S.16) can be further obtained by first rotating $\{|e_{i0}\rangle\}$ to make the norm vector of the plane where $\{|e_i\rangle\}$ are located to a specific direction $\hat{n} = (\sin\alpha\cos\beta, \sin\alpha\sin\beta, \cos\alpha)$ and then rotating them around the $\hat{n}$ direction by an angle $\gamma$, namely

$$|e_i(r_1, r_2, \alpha, \beta, \gamma)\rangle = U(\alpha, \beta, \gamma)|e_{i0}(r_1, r_2)\rangle, \quad i = 1, 2, 3, \tag{S.22}$$

where

$$U(\alpha, \beta, \gamma) = \exp\left(\frac{-i\hat{n}\cdot\vec{\sigma}\gamma}{2}\right)\exp\left(\frac{-i\sigma_z\beta}{2}\right)\exp\left(\frac{-i\sigma_y\alpha}{2}\right) \quad \text{with} \quad \hat{n} = (\sin\alpha\cos\beta, \sin\alpha\sin\beta, \cos\alpha). \tag{S.23}$$

Such an parameterization allows us to optimize the POVM $\{\Pi_i\}_{i=1}^3$ with the required target characteristic function over the parameter space of $(r_1, r_2, \alpha, \beta, \gamma)$. More specifically, based on the measurement probability of $|\psi(k_1, k_2)\rangle$ in $|e_i\rangle$, i.e.,

$$P_i(k_1, k_2; r_1, r_2, \alpha, \beta, \gamma) = |\langle\psi(k_1, k_2)|e_i(r_1, r_2, \alpha, \beta, \gamma)\rangle|^2, \tag{S.24}$$

the corresponding Fisher information matrix is given by

$$F_C(k_1, k_2; r_1, r_2, \alpha, \beta, \gamma) = \begin{pmatrix} [F_C]_{k_1k_1} & [F_C]_{k_1k_2} \\ [F_C]_{k_2k_1} & [F_C]_{k_2k_2} \end{pmatrix}, \tag{S.25}$$

with

$$[F_C]_{ij} = \sum_{k=1}^{3} \frac{1}{P_k}\frac{\partial P_k}{\partial \lambda_i}\frac{\partial P_k}{\partial \lambda_j}, \quad i, j \in \{k_1, k_2\}. \tag{S.26}$$

We numerically optimize the parameters $\{r_1, r_2, \alpha, \beta, \gamma\}$ to find the minimum of $\sqrt{\det F_C^{-1}}$ and $\text{Tr}(\mathcal{W}F_C^{-1})$.

*Three specific examples of POVM*

In the main text, we investigate three different sets of POVM and compare their performance in quantum multi-parameter estimation. The first set of POVM (which denoted as oPOVM) is given by Eq.(S.22) with the parameter values as

$$r_1 = r_2 = \sqrt{\frac{2}{3}}$$
$$\alpha = \theta, \beta = \varphi, \gamma = \pi. \tag{S.27}$$

In Fig. S.1(a), we show that $\sqrt{\det F_C^{-1}}$ associated with the oPOVM reaches the minimal value of $\sqrt{\det F_C^{-1}}$ over all 3-element rank-1 POVMs. Besides, we find that the Fisher information matrix of oPOVM is given by

$$F_C = \begin{pmatrix} 1/2 & 0 \\ 0 & \sin^2\theta/2 \end{pmatrix} = \frac{1}{2}\mathcal{F}_{S^2}. \tag{S.28}$$

In Fig. S.1(b), we demonstrate that oPOVM can also saturate the Holevo Cramér-Rao bound for the weight matrix $\mathcal{W}_1 = \mathcal{F}_{S^2}$ when estimating $\theta$ and $\varphi$. Therefore, the oPOVM is optimal for $\text{Tr}(\mathcal{W}_1\Sigma(\hat{k}_1, \hat{k}_2))$ with $\mathcal{W}_1 = J^T\mathcal{F}_{S^2}J = \mathcal{F}_{\mathbb{T}^2}$ when estimating $\mathbf{k}$. We note that the oPOVM is not optimal for another choice of the weight matrix $\mathcal{W}_2 = I$ when estimating $\theta$ and $\varphi$ (which corresponds to the weight matrix $\mathcal{W}_2 = J^TJ$ when estimating $\mathbf{k}$). Nevertheless, we can find another POVM, denoted as oPOVM($\mathcal{W}_2$), that attains the Holevo Cramér-Rao bound associated with the weight matrix $\mathcal{W}_2$, by optimizing over 3-element rank-1 POVMs, see Fig. S.1(c).

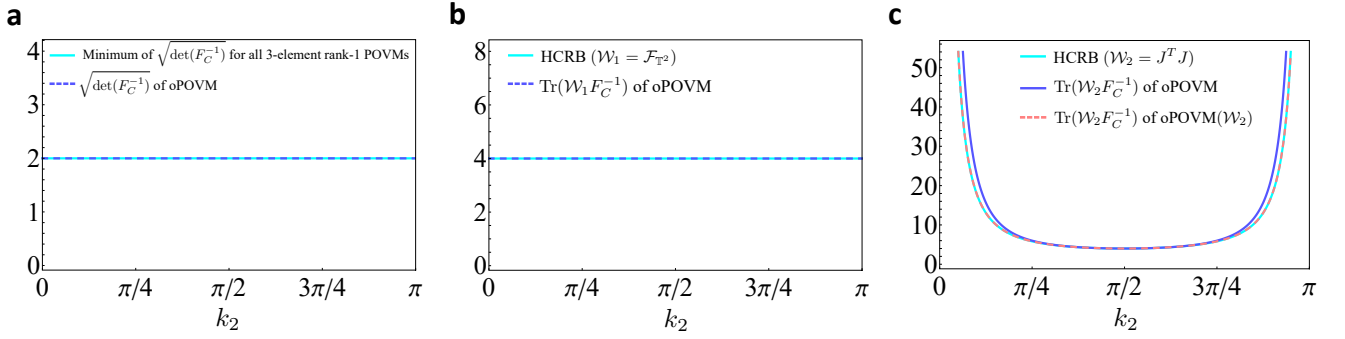

FIG. S.1. (a) $\sqrt{\det F_C^{-1}}$ associated with the oPOVM reaches the minimal value of $\sqrt{\det F_C^{-1}}$ over all 3-element rank-1 POVMs. (b) The oPOVM saturates the Holevo Cramér-Rao bound for the weight matrix $\mathcal{W}_1 = \mathcal{F}_{\mathbb{T}^2}$. (c) The oPOVM does not attain the Holevo Cramér-Rao bound for the other choice of the weight matrix $\mathcal{W}_2 = J^T J$, while the oPOVM($\mathcal{W}_2$) that is obtained by optimization saturates the Holevo Cramér-Rao bound.

The second set of POVM that we consider is a symmetric state-independent POVM (denoted as sPOVM below), which is given by Eq.(S.22) with the following parameter values

$$r_1 = r_2 = \sqrt{\frac{2}{3}},$$
$$\alpha = \pi/2, \beta = \pi/2, \gamma = \pi. \quad (S.29)$$

The corresponding state vectors $|e_i\rangle$ can be written as

$$|e_1\rangle = \sqrt{\frac{2}{3}}|0\rangle,$$
$$|e_2\rangle = -\sqrt{\frac{1}{6}}|0\rangle + \sqrt{\frac{1}{2}}|-1\rangle, \quad (S.30)$$
$$|e_3\rangle = -\sqrt{\frac{1}{6}}|0\rangle - \sqrt{\frac{1}{2}}|-1\rangle.$$

The third set of POVM in the main text is the symmetric, informationally complete POVM (SIC-POVM)[8–10] $\{|e_i\rangle\langle e_i|\}_{i=1}^4$ that has $d^2 = 4$ elements. The SIC-POVM represents the POVM with the fewest elements that can span the space of self-adjoint operators. The corresponding state vectors are given as follows

$$|e_1\rangle = \frac{1}{\sqrt{2}}|0\rangle,$$
$$|e_2\rangle = \frac{1}{\sqrt{6}}|0\rangle + \frac{1}{\sqrt{3}}|-1\rangle,$$
$$|e_3\rangle = \frac{1}{\sqrt{6}}|0\rangle + e^{i\frac{2\pi}{3}}\frac{1}{\sqrt{3}}|-1\rangle, \quad (S.31)$$
$$|e_4\rangle = \frac{1}{\sqrt{6}}|0\rangle + e^{-i\frac{2\pi}{3}}\frac{1}{\sqrt{3}}|-1\rangle.$$

**Experimental realization of POVM**

*State-dependent fluorescence measurement*

In the experiment, the state-dependent fluorescence for the realization of readout is obtained by counting the accumulated photons over many sweeps. For example, we consider to prepare the system in the spin state $|0\rangle$ and collect the signal photons of $\nu$ sweeps, where $\nu$ is usually set as a very large number. The number $n_{0i}$ of signal photons collected in the $i$th sweep is a

random variable, and we denote its expectation and variance as $\langle n_{0i} \rangle$ and $\sigma^2_{n_{0i}}$. Then the total photons collected in $\nu$ sweeps is given by

$$n_0 = \sum_{i=1}^{\nu} n_{0i}, \tag{S.32}$$

which obeys the following normal distribution according to the central limit theorem ($\nu \gg 1$),

$$n_0 \sim \mathcal{N}\left(\nu \langle n_{0i} \rangle, \nu \sigma^2_{n_{0i}}\right) \equiv \mathcal{N}(\langle n_0 \rangle, \sigma^2_{n_0}), \tag{S.33}$$

with $\langle n_0 \rangle \equiv \nu \langle n_{0i} \rangle$, $\sigma^2_{n_0} \equiv \nu \sigma^2_{n_{0i}}$. We note that similar distributions hold for the signal photons when the system is in state $|-1\rangle$ or $|1\rangle$, namely

$$\begin{aligned} n_{-1} &\sim \mathcal{N}\left(\nu \langle n_{-1i} \rangle, \nu \sigma^2_{n_{-1i}}\right) \equiv \mathcal{N}(\langle n_{-1} \rangle, \sigma^2_{n_{-1}}), \\ n_1 &\sim \mathcal{N}\left(\nu \langle n_{1i} \rangle, \nu \sigma^2_{n_{1i}}\right) \equiv \mathcal{N}(\langle n_1 \rangle, \sigma^2_{n_1}). \end{aligned} \tag{S.34}$$

*Projective measurement in an extended Hilbert space*

In our experiments, the system is encoded in the Hilbert space spanned by $\{|0\rangle, |-1\rangle\}$. We realize the 3-element POVM on the system via the projective measurement in the extended three-level Hilbert space $\{|0\rangle, |-1\rangle, |+1\rangle\}$. First, we choose the projective basis states $|u_i\rangle$ such that

$$(|0\rangle\langle 0| + |-1\rangle\langle -1|) |u_i\rangle = |e_i\rangle. \tag{S.35}$$

For $|e_i\rangle$ in Eq.(3) of the main text, the explicit form of $|u_i\rangle = \langle 0|u_i\rangle |0\rangle + \langle -1|u_i\rangle |-1\rangle + \langle +1|u_i\rangle |+1\rangle$ is given by

$$\begin{aligned} |u_1\rangle &= r_1 \cos\frac{\theta_1}{2} |0\rangle + r_1 \sin\frac{\theta_1}{2} e^{i\varphi_1} |-1\rangle + \sqrt{1-r_1^2} |+1\rangle, & (S.36) \\ |u_2\rangle &= r_2 \cos\frac{\theta_2}{2} |0\rangle + r_1 \sin\frac{\theta_2}{2} e^{i\varphi_2} |-1\rangle - \sqrt{1-r_2^2} e^{i \arg\langle e_1|e_2\rangle} |+1\rangle, & (S.37) \\ |u_3\rangle &= r_3 \cos\frac{\theta_3}{2} |0\rangle + r_3 \sin\frac{\theta_3}{2} e^{i\varphi_3} |-1\rangle - \sqrt{1-r_3^2} e^{i \arg\langle e_1|e_3\rangle} |+1\rangle. & (S.38) \end{aligned}$$

By such a choice, we have

$$\mathrm{Tr}\left(\Pi_i \rho_{\boldsymbol{k}}\right) = \mathrm{Tr}\left(|e_i\rangle\langle e_i| \rho_{\boldsymbol{k}}\right) = \langle u_i | \rho_{\boldsymbol{k}} | u_i \rangle \equiv p_i \tag{S.39}$$

In order to realize the projective measurement $\{|u_i\rangle\langle u_i|\}$, we apply a unitary rotation $U_1$ on the NV center spin, which rotates the system by the following transformation

$$\begin{aligned} U_1 |u_1\rangle &= |+1\rangle, \\ U_1 |u_2\rangle &= |0\rangle, \\ U_1 |u_3\rangle &= |-1\rangle. \end{aligned} \tag{S.40}$$

The subsequent state-dependent fluorescence measurement provides the number of accumulated photons $n_j(1)$. Similarly, we apply another unitary rotation $U_2$ on the NV center spin, which rotates the system by the following transformation

$$\begin{aligned} U_2 |u_1\rangle &= |+1\rangle, \\ U_2 |u_2\rangle &= |-1\rangle, \\ U_2 |u_3\rangle &= |0\rangle. \end{aligned} \tag{S.41}$$

The corresponding number of accumulated photons is denoted as $n_j(2)$. The unitary rotations are implemented by engineering microwave driving fields on resonance with both transitions ($0 \leftrightarrow -1$ and $0 \leftrightarrow +1$), see Fig.S.2. The corresponding driven Hamiltonians are written as

$$\begin{aligned} H_d^{(1)} &= H_0 + \Omega_1 \cos(\omega_1 t + \beta) \left(|0\rangle\langle -1| + |-1\rangle\langle 0|\right), & t \in [t_1, t_1 + \tau_1] & (S.42) \\ H_d^{(2)} &= H_0 + \Omega_2 \cos(\omega_2 t + \beta) \left(|0\rangle\langle +1| + |+1\rangle\langle 0|\right), & t \in [t_2, t_2 + \tau_2] & (S.43) \end{aligned}$$

with $H_0 = \omega_1 |-1\rangle \langle -1| + \omega_2 |+1\rangle \langle +1|$, where $\omega_{1,2}$ denote the energy gaps between the states $|0\rangle, |-1\rangle$ and $|0\rangle, |+1\rangle$ respectively, and $\tau_{1,2}$ represent the time duration of individual microwave-field pulses. With different values of the phase $\beta$, the above driven Hamiltonians are able to achieve arbitrary rotations in the subspaces spanned by $\{|0\rangle, |-1\rangle\}$ and $\{|0\rangle, |+1\rangle\}$ respectively. In order to implement the unitary rotation $U_1$, we first rotate the projective part of $|u_1\rangle$ in the subspace of $\{|0\rangle, |-1\rangle\}$ to the state $|0\rangle$ by invoking $H_d^{(1)}$. Then, the obtained superposition state of $|0\rangle$ and $|+1\rangle$ is further rotated to the state $|+1\rangle$ through the second driven Hamiltonian $H_d^{(2)}$. In this process, $|u_2\rangle$ and $|u_3\rangle$ are rotated to the subspace of $\{|0\rangle, |-1\rangle\}$. Finally, we again apply $H_d^{(1)}$ to rotate $|u_2\rangle$ and $|u_3\rangle$ to the state $|0\rangle$ and $|-1\rangle$. Hence, we realize a unitary transformation in Eq. (S.40). We note that the other unitary transformation $U_2$ can be implemented in a similar way.

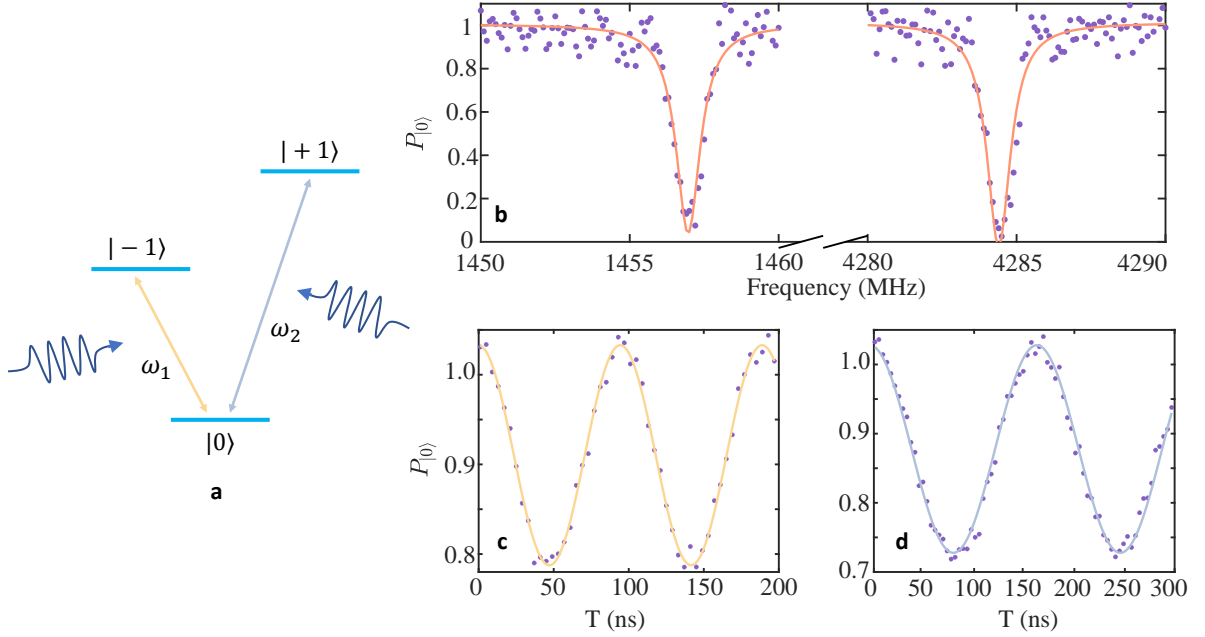

FIG. S.2. (a) The energy level structure of the NV center spin in diamond under an external magnetic field and microwave driving. (b) We measure the energy gap $\omega_{1,2}$ through pulsed optically detected magnetic resonance (Pulsed-ODMR). (c-d) Rabi oscillations between the states $|0\rangle, |-1\rangle$ and $|0\rangle, |+1\rangle$, respectively.

After applying the unitary transformations $U_1$ and $U_2$ [see Eq.(S.40) and Eq.(S.41)], using Eq.(S.33-S.34) and the Lyapunov central limit theorem, the number of signal photons collected in the total $\nu$ sweeps will be

$$n_j(1) \sim \mathcal{N}\left(p_1 \langle n_1\rangle + p_2 \langle n_0\rangle + p_3 \langle n_{-1}\rangle, \ p_1\sigma_{n_1}^2 + p_2\sigma_{n_0}^2 + P_3\sigma_{n_{-1}}^2\right), \tag{S.44}$$

$$n_j(2) \sim \mathcal{N}\left(p_1 \langle n_1\rangle + p_2 \langle n_{-1}\rangle + p_3 \langle n_0\rangle, \ p_1\sigma_{n_1}^2 + p_2\sigma_{n_{-1}}^2 + p_3\sigma_{n_0}^2\right). \tag{S.45}$$

Based on the obtained photon number $n_j(1)$ and $n_j(2)$, we can get the following probability associated with the $j$-th measurement as

$$\begin{aligned} p_{1j} &= \frac{\langle n_0\rangle + \langle n_{-1}\rangle - n_j(1) - n_j(2)}{\langle n_0\rangle + \langle n_{-1}\rangle - 2\langle n_1\rangle}, \\ p_{2j} &= \frac{(\langle n_0\rangle - \langle n_1\rangle)n_j(1) + (\langle n_1\rangle - \langle n_{-1}\rangle)n_j(2) - (\langle n_0\rangle - \langle n_{-1}\rangle)\langle n_1\rangle}{(\langle n_0\rangle - \langle n_{-1}\rangle)(\langle n_0\rangle + \langle n_{-1}\rangle - 2\langle n_1\rangle)}, \\ p_{3j} &= 1 - p_{1j} - p_{2j}, \end{aligned} \tag{S.46}$$

which satisfy the following equations

$$\begin{aligned} p_{1j} + p_{2j} + p_{3j} &= 1 \\ p_{1j}\langle n_1\rangle + p_{2j}\langle n_0\rangle + p_{3j}\langle n_{-1}\rangle &= n_j(1) \\ p_{1j}\langle n_1\rangle + p_{2j}\langle n_{-1}\rangle + p_{3j}\langle n_0\rangle &= n_j(2). \end{aligned} \tag{S.47}$$

Note that for two independent random variables $\xi \sim \mathcal{N}(\mu_1, \sigma_1^2)$ and $\eta \sim \mathcal{N}(\mu_2, \sigma_2^2)$, $a\xi + b\eta \sim \mathcal{N}(a\mu_1 + b\mu_2, a^2\sigma_1^2 + b^2\sigma_2^2)$

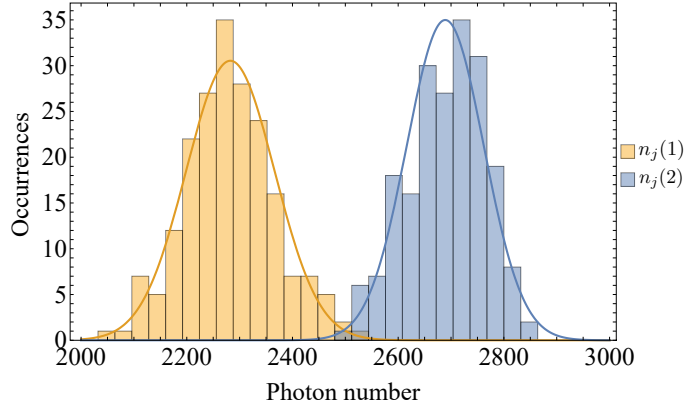

FIG. S.3. An example of the histogram of occurrences of $n_j(1)$ and $n_j(2)$ in the experiment.

holds. Hence, $p_{1j}$ and $p_{2j}$ obey normal distributions as well, namely

$$\begin{aligned} p_{1j} &\sim \mathcal{N}\left(p_1, \tilde{\sigma}_1^2\right), \\ p_{2j} &\sim \mathcal{N}\left(p_2, \tilde{\sigma}_2^2\right), \\ p_{3j} &\sim \mathcal{N}\left(p_3, \tilde{\sigma}_3^2\right), \end{aligned} \tag{S.48}$$

with the variance $\tilde{\sigma}_i^2$ ($i = 1, 2, 3$) given by

$$\begin{aligned} \tilde{\sigma}_1^2 &= \frac{2p_1\sigma_{n_1}^2 + (p_2 + p_3)(\sigma_{n_0}^2 + \sigma_{n_{-1}}^2)}{(\langle n_0 \rangle + \langle n_{-1} \rangle - 2\langle n_1 \rangle)^2} \\ \tilde{\sigma}_2^2 &= \frac{(\langle n_0 \rangle - \langle n_1 \rangle)^2 (p_1\sigma_{n_1}^2 + p_2\sigma_{n_0}^2 + p_3\sigma_{n_{-1}}^2) + (\langle n_1 \rangle - \langle n_{-1} \rangle)^2 (p_1\sigma_{n_1}^2 + p_2\sigma_{n_{-1}}^2 + p_3\sigma_{n_0}^2)}{(\langle n_0 \rangle - \langle n_{-1} \rangle)^2 (\langle n_0 \rangle + \langle n_{-1} \rangle - 2\langle n_1 \rangle)^2}, \\ \tilde{\sigma}_3^2 &= \frac{(\langle n_1 \rangle - \langle n_{-1} \rangle)^2 (p_1\sigma_{n_1}^2 + p_2\sigma_{n_0}^2 + p_3\sigma_{n_{-1}}^2) + (\langle n_0 \rangle - \langle n_1 \rangle)^2 (p_1\sigma_{n_1}^2 + p_2\sigma_{n_{-1}}^2 + p_3\sigma_{n_0}^2)}{(\langle n_0 \rangle - \langle n_{-1} \rangle)^2 (\langle n_0 \rangle + \langle n_{-1} \rangle - 2\langle n_1 \rangle)^2}. \end{aligned} \tag{S.49}$$

We remark that these variance are inversely proportional to the number of sweeps $\nu$, i.e. $\tilde{\sigma}_1^2 \propto \nu^{-1}$, $\tilde{\sigma}_2^2 \propto \nu^{-1}$, which indicates that the fluctuation of $p_{1j}$ and $p_{2j}$ can be reduced by increasing the number of sweeps in the $j$th-experiment. Here, we assume that $\tilde{\sigma}_i^2$ is small enough to ensure that $p_{1j} > 0$ and $p_{2j} > 0$. Therefore, according to Eq.(S.39), we can obtain the measurement probabilities for the POVM $\{\Pi = |e_i\rangle\langle e_i|\}_{i=1}^3$, namely

$$\text{Tr}\left(\Pi_i \rho_{\boldsymbol{k}}\right) = \langle u_i | \rho_{\boldsymbol{k}} | u_i \rangle = p_i = \langle p_{ij} \rangle. \tag{S.50}$$

We can then assign $x_j \in \{1, 0, -1\}$ as the measurement outcome of the $j$-th experimental according to the obtained probabilities $p_{1j}, p_{2j}$ and $p_{3j}$ respectively [1].

### Experimental procedure to extract the covariance matrix

The full covariance matrix $\Sigma(\hat{\boldsymbol{k}})$ contains not only the measurement uncertainties for the $j$-th parameter $\langle \delta \hat{\boldsymbol{k}}_j^2 \rangle$, but also the cross elements $\langle \delta \hat{\boldsymbol{k}}_i \delta \hat{\boldsymbol{k}}_j \rangle$. In this section, we will present details on how to experimentally extract the covariance matrix $\Sigma(\hat{k}_1, \hat{k}_2)$ and thereby directly determine the performance of quantum multi-parameter estimation protocols.

*Maximum likelihood estimator*

We consider the general POVM $\{\Pi_i\}_{i=1}^m$ with $m$ elements. The measurement outcome corresponding to the $i$-th measurement $\Pi_i$ is denoted as $a_i$, with the probability $p_i(\boldsymbol{k}) = \text{Tr}[\Pi_i \rho_{\boldsymbol{k}}]$. By performing $N$ repetitive measurements, we obtain a series

of measurement outcomes as $\vec{x} = (x_1, x_2, \cdots, x_j, \cdots, x_N)$, where $x_j \in (a_1, \cdots, a_m)$ associated with the probability $P_j(x_j, \boldsymbol{k}) \in \{p_1(\boldsymbol{k}), \cdots, p_m(\boldsymbol{k})\}$. Next, we present the detailed steps to construct the maximum likelihood estimator (MLE) $\hat{\boldsymbol{k}}$ based on $\vec{x}$. The log-likelihood function is

$$\ell(\boldsymbol{k}; \vec{x}) = \sum_{j=1}^{N} \ln P_j(x_j, \boldsymbol{k}). \tag{S.51}$$

The MLE $\hat{\boldsymbol{k}}$ maximizes $\ell(\boldsymbol{k}; \vec{x})$ and thus satisfies the likelihood equations as $\partial \ell / \partial k_1 |_{\boldsymbol{k} = \hat{\boldsymbol{k}}} = 0$, $\partial \ell / \partial k_2 |_{\boldsymbol{k} = \hat{\boldsymbol{k}}} = 0$, namely

$$\begin{aligned}
\sum_{j=1}^{N} \frac{1}{P_j(x_j, \hat{\boldsymbol{k}})} \frac{\partial P_j(x_j, \hat{\boldsymbol{k}})}{\partial \hat{k}_1} &= \sum_{i=1}^{m} \frac{n_i}{p_i(\hat{\boldsymbol{k}})} \frac{\partial p_i(\hat{\boldsymbol{k}})}{\partial \hat{k}_1} = 0, \\
\sum_{j=1}^{N} \frac{1}{P_j(x_j, \hat{\boldsymbol{k}})} \frac{\partial P_j(x_j, \hat{\boldsymbol{k}})}{\partial \hat{k}_2} &= \sum_{i=1}^{m} \frac{n_i}{p_i(\hat{\boldsymbol{k}})} \frac{\partial p_i(\hat{\boldsymbol{k}})}{\partial \hat{k}_2} = 0,
\end{aligned} \tag{S.52}$$

where $n_i$ is the number of occurrences of $a_i$ in $\vec{x}$. We construct the probability estimator of $p_i(\boldsymbol{k})$ as follows

$$\hat{p}_i(\vec{x}) = \frac{1}{N} \sum_{j=1}^{N} \delta_{a_i, x_j} \equiv \frac{1}{N} \sum_{j=1}^{N} y_i^{(j)} = \frac{n_i}{N}. \tag{S.53}$$

According to Eq. (S.52), we can get

$$\begin{aligned}
\sum_{i=1}^{m} \frac{\hat{p}_i(\vec{x})}{p_i(\hat{\boldsymbol{k}})} \frac{\partial p_i(\hat{\boldsymbol{k}})}{\partial \hat{k}_1} &= 0, \\
\sum_{i=1}^{m} \frac{\hat{p}_i(\vec{x})}{p_i(\hat{\boldsymbol{k}})} \frac{\partial p_i(\hat{\boldsymbol{k}})}{\partial \hat{k}_2} &= 0.
\end{aligned} \tag{S.54}$$

Using the normalization condition $\sum_i^m \hat{p}_i = 1$, the MLE $\hat{\boldsymbol{k}}$ as functions of $\{\hat{p}_i(\vec{x})\}_{i=1}^{m-1}$ can be solved based on Eq. (S.54),

$$\hat{\boldsymbol{k}}(\vec{x}) = \hat{\boldsymbol{k}}\left(\hat{p}_1(\vec{x}), \hat{p}_2(\vec{x}), \cdots, \hat{p}_{m-1}(\vec{x})\right). \tag{S.55}$$

It is generally a challenging task to get the analytical expression of $\hat{\boldsymbol{k}}(\hat{p}_1, \hat{p}_2, \cdots, \hat{p}_{m-1})$, thus one usually needs to solve Eq. (S.54) numerically in order to get the maximum likelihood estimator.

We note that when the number of POVM elements $m$ is equal to to the number of parameters plus one, the maximum likelihood estimator have a simple form. In the experiment, we use the set of 3-element POVM. In such a scenario, we have

$$\begin{aligned}
p_1 &= p_1(k_1, k_2) \\
p_2 &= p_2(k_1, k_2) \\
p_3 &= p_3(k_1, k_2) = 1 - p_1(k_1, k_2) - p_2(k_1, k_2).
\end{aligned} \tag{S.56}$$

Under the assumption that $(p_1, p_2)$ and $(k_1, k_2)$ are one-to-one mapping, we can solve $k_1$ and $k_2$ as functions of $p_1$ and $p_2$ as

$$\begin{aligned}
k_1 &= k_1(p_1, p_2), \\
k_2 &= k_2(p_1, p_2).
\end{aligned} \tag{S.57}$$

Then, one can prove that the following estimators

$$\begin{aligned}
\hat{k}_1 &= k_1(\hat{p}_1, \hat{p}_2), \\
\hat{k}_2 &= k_2(\hat{p}_1, \hat{p}_2),
\end{aligned} \tag{S.58}$$

satisfy the likelihood equations and thus yield the MLE $\hat{\boldsymbol{k}}$.

*Extracting the covariance matrix from measurements*

By exploiting the one-to-one map between $\boldsymbol{k} = (k_1, k_2) \leftrightarrow \boldsymbol{n_k} = (\theta_{\boldsymbol{k}}, \varphi_{\boldsymbol{k}})$ in a suitable parameter range, the covariance matrix of the MLE in Eq. (S.58) can be expressed as

$$\Sigma(\hat{k}_1, \hat{k}_2) = \left[\frac{\partial(\hat{k}_1, \hat{k}_2)}{\partial(\hat{p}_1, \hat{p}_2)}\right] \Sigma(\hat{p}_1, \hat{p}_2) \left[\frac{\partial(\hat{k}_1, \hat{k}_2)}{\partial(\hat{p}_1, \hat{p}_2)}\right]^T$$

$$= J^{-1} \left[\frac{\partial(\hat{\theta}, \hat{\varphi})}{\partial(\hat{p}_1, \hat{p}_2)}\right] \Sigma(\hat{p}_1, \hat{p}_2) \left[\frac{\partial(\hat{\theta}, \hat{\varphi})}{\partial(\hat{p}_1, \hat{p}_2)}\right]^T (J^{-1})^T \quad (S.59)$$

where $J$ is the Jacobian matrix written as follows

$$J = \frac{\partial(\theta, \varphi)}{\partial(k_1, k_2)}. \quad (S.60)$$

Note that for the MLE of 3-element POVM, $\partial(\hat{\theta}, \hat{\varphi})/\partial(\hat{p}_1, \hat{p}_2)$ and $\partial(\theta, \varphi)/\partial(p_1, p_2)$ have the same functional form, therefore we can obtain $\partial(\hat{\theta}, \hat{\varphi})/\partial(\hat{p}_1, \hat{p}_2)$ from the slope of the signals, see Fig. S.4.

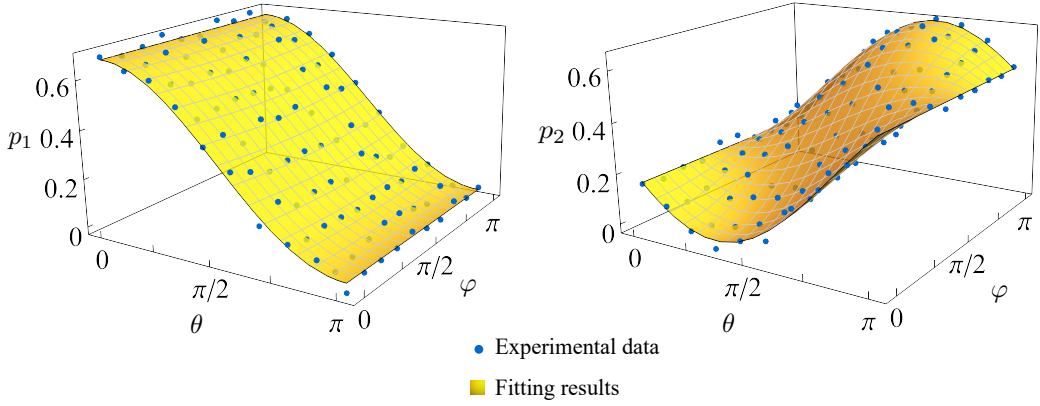

FIG. S.4. Examples of the experimental measurement of $\partial(\hat{\theta}, \hat{\varphi})/\partial(\hat{p}_1, \hat{p}_2)$. From Eq. (S.46) and (S.48), we can estimate $p_1(\theta, \varphi)$ and $p_2(\theta, \varphi)$ in the experiment. By measuring $p_1, p_2$ at different $(\theta, \varphi)$, we derive the fitting functions $p_1(\theta, \varphi)$ and $p_2(\theta, \varphi)$. Based on the fitting results, we can obtain the derivative components in the Jacobian matrix. (a) Measurement of $\partial p_1/\partial \theta$ and $\partial p_1/\partial \varphi$ by fitting $p_1(\theta, \varphi)$. (b) Measurement of $\partial p_2/\partial \theta$ and $\partial p_2/\partial \varphi$ by fitting $p_2(\theta, \varphi)$.

Note that the definition for $\{\hat{p}_1, \hat{p}_2\}$ in Eq.(S.53) can be re-expressed as

$$\hat{p}_i(\vec{x}) = \frac{1}{N} \sum_{j=1}^{N} y_i^{(j)}, \quad (S.61)$$

with $y_i^{(j)} \equiv \delta_{a_i, x_j}$, i.e.

$$y_1^{(j)} = \delta_{a_1, x_j} = \delta_{1, x_j} = \frac{x_j + |x_j|}{2},$$
$$y_2^{(j)} = \delta_{a_2, x_j} = \delta_{0, x_j} = 1 - |x_j|, \quad (S.62)$$

where we have set the measurement outcomes as $a_1 = 1$, $a_2 = 0$ and $a_3 = -1$ respectively. One can see that $\hat{p}_i(\vec{x})$ can be viewed as the average of $N$ independent identical measurements. By defining $y_i \equiv \delta_{a_i, x}$ with $x$ the outcome of single measurement, we have

$$\Sigma(\hat{p}_1, \hat{p}_2) = \frac{1}{N} \Sigma(y_1, y_2). \quad (S.63)$$

The right-hand side of the above equation, i.e. $\Sigma(y_1, y_2)$, can be extracted from the experimental result $\{y_i^{(1)}, \cdots, y_i^{(N)}\}$ via the well-known Bessel's formula as

$$\Sigma(y_1, y_2)_{ij} \equiv \mathrm{Cov}(y_i, y_j) = \frac{1}{N-1} \sum_{k=1}^{N} (y_i^{(k)} - \bar{y}_i)(y_j^{(k)} - \bar{y}_j), \tag{S.64}$$

where

$$\bar{y}_i = \frac{1}{N} \sum_{k=1}^{N} y_i^{(k)}. \tag{S.65}$$

Therefore, we can experimentally extract $\Sigma(\hat{k}_1, \hat{k}_2)$ from $\partial(\hat{\theta}, \hat{\varphi})/\partial(\hat{p}_1, \hat{p}_2)$ and the covariance matrix $\Sigma(\hat{p}_1, \hat{p}_2)$.

**Metrological characterization of topological bands**

*Topological invariant, quantum volume and metrological potential*

In this section, we briefly review the relation between the topological invariants, the quantum volume and the bound for quantum metrology, which is related to our experiments. We refer further details to Ref.[3]. It is important to note that the left-hand side of the metric-curvature correspondence relation in Eq. (S.6) integrates to an integer topological invariant of the occupied Bloch band,

$$\mathrm{Ch}_n = \frac{1}{n!} \left(\frac{i}{2\pi}\right)^n \int_{\mathbb{T}^{2n}} \mathrm{Tr}(\Omega^n) \in \mathbb{Z}, \tag{S.66}$$

which is known as the $n$-th Chern number. This topological invariant is central in the classification of Chern insulators. The metric-curvature correspondence relation in Eq. (S.6) implies the inequality (see Eq.(18) in Ref. [3])

$$|\mathrm{Ch}_n| \leqslant \frac{(2n)!}{2^{n(n-1)+1} n! \pi^n} \mathrm{vol}_g(\mathbb{T}^{2n}), \tag{S.67}$$

where the *quantum volume* (also known as the complexity of the band) is defined as [3]

$$\mathrm{vol}_g(\mathbb{T}^{2n}) = \int_{\mathbb{T}^{2n}} \sqrt{\det(g)} \mathrm{d}^{2n} k. \tag{S.68}$$

We note that the equality in the above equation is satisfied provided $\mathrm{sgn}(d\vec{n})$ is constant (everywhere where it is meaningful, i.e., where $\sqrt{\det(g)} \neq 0$), or, equivalently, if the function on the left-hand side of Eq. (S.6) does not change sign. We can see that quantum volume is an important quantity related to quantum topology from Eq. (S.67).

From the quantum SLD-CRB in Eq. (S.4), we can establish a relationship between the performance of quantum parameter estimation and the quantum volume as [3]

$$\int_{\mathbb{T}^d} \left(\sqrt{\det \Sigma(\hat{\mathbf{k}})}\right)^{-1} \mathrm{d}\boldsymbol{k} \leqslant 2^d N^{d/2} \mathrm{vol}_g(\mathbb{T}^d), \tag{S.69}$$

where the left-hand side is a quantity that characterizes the performance of global sensing [11]. Furthermore, in the case of Eq. (S.67) taking the equal sign, we can establish a fundamental connection between quantum parameter estimation and topological invariants

$$\int_{\mathbb{T}^{2n}} \left(\sqrt{\det \Sigma(\hat{\mathbf{k}})}\right)^{-1} \mathrm{d}\mathbf{k} \leqslant N^{d/2} \frac{2^{n(n+1)+1} n! \pi^n}{(2n)!} |\mathrm{Ch}_n|. \tag{S.70}$$

In the two-dimensional case, the above equation becomes

$$\int_{\mathbb{T}^2} \left(\sqrt{\det \Sigma(\hat{\mathbf{k}})}\right)^{-1} \mathrm{d}k_1 \mathrm{d}k_2 \leqslant 4N \mathrm{vol}_g(\mathbb{T}^2) = 2N \int_{\mathbb{T}^2} |\Omega_{12}(\mathbf{k})| \mathrm{d}k_1 \mathrm{d}k_2. \tag{S.71}$$

And if Eq. (S.67) takes the equal sign, which is the case for the model in Eq. (9) that we study in the main text, we can get

$$\int_{\mathbb{T}^2} \left(\sqrt{\det \Sigma(\hat{\mathbf{k}})}\right)^{-1} dk_1 dk_2 \leqslant 4\pi N \left|\text{Ch}_1\right|. \tag{S.72}$$

The above discussions focus on the systems that has no chiral symmetry, while similar conclusions hold for the systems with chiral symmetry. We now set $d = D - 1 = 2n - 1$ for some integer $n > 0$. In this case, $H(\mathbf{k})$ is also generically gapped but chiral symmetry is now present. In this case, the topological invariant of the occupied Bloch band is the winding number $\nu$, which completely classifies the topological phase

$$\nu = (-1)^{n-1} \left(\frac{i}{2\pi}\right)^n \frac{(n-1)!}{(2n-1)!} \int_{\mathbb{T}^{2n}} \text{Tr}\left[\left(q^{-1}dq\right)^{2n-1}\right] \in \mathbb{Z}. \tag{S.73}$$

Similar to Eq. (S.67), we have an inequality for the winding number (see Eq.(37) in Ref. [3])

$$|\nu| \leqslant \frac{(n-1)!}{2^{\frac{1}{2}(n-1)(2n-5)}\pi^n} \text{vol}_g\left(\mathbb{T}^{2n-1}\right). \tag{S.74}$$

Again, we note that the equality is satisfied when $\text{sgn}(d\vec{n})$ is constant (for $\sqrt{\det(g)} \neq 0$)). In this case, we can get an equation similar to Eq. (S.70) when Eq. (S.74) takes the equal sign

$$\int_{\mathbb{T}^{2n-1}} \left(\sqrt{\det \Sigma(\hat{\mathbf{k}})}\right)^{-1} d^{2n-1}k \leqslant N^{d/2} \frac{2^{\frac{1}{2}(n-1)(2n-1)+1}\pi^n}{(n-1)!} |\nu|. \tag{S.75}$$

### Independent experimental measurement of quantum geometric tensor

The QGT data for Fig.2 and Fig.3 in the main text are extracted by measuring the fidelity between neighbouring quantum state in parameter space [12]. The fidelity between neighboring quantum states in parameter space is given by

$$|\langle\psi(\mathbf{k}+d\mathbf{k})|\psi(\mathbf{k})\rangle|^2 = 1 - \sum_{i,j} g_{ij}(\mathbf{k}) dk_i dk_j + O(dk_i dk_j dk_l), \tag{S.76}$$

where $g_{ij}(\mathbf{k})$ is the quantum metric in momentum space. So we can experimentally prepare the state $|\psi(\mathbf{k}+d\mathbf{k})\rangle$ and measure the overlap between $|\psi(\mathbf{k}+d\mathbf{k})\rangle$ and $|\psi(\mathbf{k})\rangle$. By changing $d\mathbf{k}$ in a range that $O(dk_i dk_j dk_l)$ is negligible and repeat the above measurement, we can extract the quantum metric by fitting to the function in Eq.(S.76). For Dirac systems, due to the metric-curvature correspondence, we can further obtain the Berry curvature from the measurement results of the quantum metric. Therefore, we can experimentally extract the complete quantum geometry tensor using the techniques as we developed in Ref. [12]. The Berry curvature data for Fig.4 in the main text is from Ref. [13], where we extract the complete QGT by weak modulations of the parameters.

[1] M. Yu, Y. Liu, P. Yang, M. Gong, Q. Cao, S. Zhang, H. Liu, M. Heyl, T. Ozawa, N. Goldman, and J. Cai, Quantum Fisher information measurement and verification of the quantum Cramér–Rao bound in a solid-state qubit, npj Quantum Inf. **8**, 56 (2022).

[2] J. Liu, H. Yuan, X.-M. Lu, and X. Wang, Quantum Fisher information matrix and multiparameter estimation, J. Phys. A Math. Theor. **53**, 023001 (2020).

[3] B. Mera, A. Zhang, and N. Goldman, Relating the topology of Dirac Hamiltonians to quantum geometry: When the quantum metric dictates Chern numbers and winding numbers, SciPost Phys. **12**, 18 (2022).

[4] R. Demkowicz-Dobrzański, W. Górecki, and M. Guţă, Multi-parameter estimation beyond quantum Fisher information, J. Phys. A Math. Theor. **53**, 363001 (2020).

[5] F. Albarelli, M. Barbieri, M. Genoni, and I. Gianani, A perspective on multiparameter quantum metrology: From theoretical tools to applications in quantum imaging, Phys. Lett. A **384**, 126311 (2020).

[6] A. Carollo, B. Spagnolo, A. A. Dubkov, and D. Valenti, On quantumness in multi-parameter quantum estimation, Journal of Statistical Mechanics: Theory and Experiment **2019**, 094010 (2019).

[7] C. Li, M. Chen, and P. Cappellaro, A geometric perspective: experimental evaluation of the quantum Cramer-Rao bound, arXiv:2204.13777 (2022).

[8] J. M. Renes, R. Blume-Kohout, A. J. Scott, and C. M. Caves, Symmetric informationally complete quantum measurements, J. Math. Phys. **45**, 2171 (2004).



\end{document}